# Optimal number of users in Co-operative spectrum sensing in WRAN using Cyclo-Stationary Detector

Manish B Dave

*Assistant Professor, C. U. Shah College of Engg. & Tech., Wadhwan city-363030, Surendranagar Gujarat India*

*Abstract*— **Cognitive radio allows unlicensed users to access licensed frequency bands through dynamic spectrum access so as to reduce spectrum scarcity. This requires intelligent spectrum sensing techniques. This paper investigates the use of cyclo-stationary detector and performance evaluation for Digital Video Broadcast-Terrestrial (DVB-T) signals. Generally, DVB-T is specified in IEEE 802.22 standard in VHF and UHF TV broadcasting spectrum. Simulations results show that implementing co-operative spectrum sensing help in better utilization of resources. The paper further proposes to find number of optimal users in a scenario to optimize the detection probability and makes use of the particle swarm optimization (PSO) technique to find an optimum value of threshold.**

*Keywords*-**Cognitve Radio, WRAN, Spectrum Sensing, Co-operative Spectrum Sensing, Cyclo-Stationary detector, DVB-T, Particle Swarm Optimization(PSO), Threshold Adaptation.**

## I. INTRODUCTION

Cognitive radio is a concept that is still in its nascent stage. It is based on the concept of dynamic frequency allocation which is in contrast to the statically allocated frequencies which is prevalent throughout the world at present time. Being a dynamic frequency concept it never relies on a single frequency, it continuously looks for frequency which are free to be used and as a result solves the issue of scarcity of spectrum. The cognitive radio makes use of the fact provided by the FCC's Spectrum Policy Task Force that occupation of frequency bands by users at any instant of time varies from 15% to 85%. This usage variation has led to the genesis of cognitive radio as researchers have found it a much viable alternative to reassigning the frequency spectrum [1]. Cognitive radios present a much needed solution so that the underutilized bands can be opportunistically used in such a manner that the licensed users are not affected in their communication. Thus one can define the cognitive radio as a system that senses its operating environment i.e. the frequency spectrum and depending on the measurements it automatically adjusts it operating parameters like frequency band of operation, thus ensuring fast communication and minimal interference to the licensed users.

The broadcast TV's frequency spectrum comes out to be the foremost choice for cognitive radio as it has comparatively low propagation loss as compared to other frequencies in addition to its varying usage patterns.

IEEE 802.22 standard or Wireless Regional Area Network (WRAN) is the first standard based on Cognitive Radio. The standard's operation frequency ranges from 54 MHz to 862 MHz i.e. VHF and UHF portion of the frequency spectrum. The licensed users in these bands include North America's Advanced Television System Committee (ATSC) DTV having a bandwidth of 6 MHz, Europe's Digital Video Broadcast-Terrestrial (DVB-T) having a bandwidth of 8 MHz and microphone users employing FM with 200 kHz bandwidth [2].

The paper is organized in 6 sections. Following this introduction section, spectrum sensing and its application in cognitive radios is presented in section II. Cyclo-stationary method of spectrum sensing and its application in WRAN are highlighted in section III whereas section IV presents an algorithm for finding optimum number of users in case of co-operative spectrum sensing. An algorithm for applying the particle swarm optimization for threshold adaptation is illustrated in section VI. Section VI presents the simulation results followed by conclusion in section VII.

The paper proposes to find out minimal number of users required for satisfactory detection performance in case of co-operative sensing. This results in minimum overhead and enables fast response. Further it proposes a method of applying the PSO technique to cognitive radio threshold adaptation.

## II. SPECTRUM SENSING

The important characteristics of the cognitive radio [3] can be enumerated as (1) sensing the spectrum and determining empty bands, (2) managing the spectrum by selecting the best available spectrum channels in case of multiple options, (3) allow spectrum sharing, and (4) provide spectrum mobility whereby secondary users vacate the band as soon as a primary users enters the band and switch to another empty frequency.

Spectrum sensing turns out to be the first and a very important part of the cognitive radio system. It is simply checking a particular frequency band for the presence of a primary or licensed user. Analytically the spectrum sensing can be reduced to a simple identification problem [4], formalized as a hypothesis test.

$$H1 : x(n) = s(n) + w(n)$$
$$H0 : x(n) = w(n)$$

where x(n) is the signal received by secondary users, s(n) is the signal transmitted by primary user, and w(n) is additive white Gaussian noise with variance $\sigma^2_w$.

H0 indicates that the frequency band only has noise and there is no presence of primary user. H1 denotes that a primary user is present in the band that is being sensed. Selecting H0 or H1 is done by comparing the value of 'x(n)' with a pre-defined threshold value 'λ'. Taking into account the energy of a signal here,

$$E = \sum |x(n)|^2$$

the condition for the two hypothesis can be defined as below:-

$$1) H1 : E > \lambda$$
$$2) H0 : E < \lambda$$

Probability of detection ($P_d$) meaning H1 to be true with H1 being the correct state. Probability of false alarm ($P_f$) being H1 coming out to be true when it should have been H0. In case of H0 being true when it should have been H1 we get the probability of miss-detection ($P_m$).

There are number of different spectrum sensing methods for determining vacant bands. Most common of them is the energy detector technique wherein energy is computed to check for primary users. Its main advantages are simple implementation and low computation complexity.

Matched filter sensing technique presents an optimal way for detection by maximizing the SNR and also takes less time. But it requires prior knowledge of the signal and different types of receivers for different types of signal types and thus complexity increases.

Cyclo-stationary detector utilizes the periodicities in the signal to detect the presence of primary users in the bands. Although complexity is more the energy detector but it works for low SNR conditions unlike the energy detector. There is no requirement for different types of receivers as in case of matched filter detector.

In order to enhance the detection probability co-operative sensing scheme [5]- [6] is adopted wherein numbers of different users report their decision regarding presence of primary rules. Global decision is based on the fusion rule; the most common of them is k out of n rule. It is given as

$$P_x = \sum_{i=k}^{n} \binom{n}{i} p_x^i (1-p_x)^{n-i}$$

where     k=n,        AND operation
           k=1,        OR operation
           k=ceil(n/2),     MAJORITY operation

Here '$P_x$' being the probability of detection and '$p_x$' is the probability of detection for single users.

### III. CYCLO-STATIONARY ENERGY DETECTOR

A number of different schemes are used for detecting the presence of signals in a given frequency band. Some of the approaches use the signal energy or some specific characteristics of the signal to identify its presence and even its type. This section outlines the Cyclo-Stationary detector, a signal specific detector, and its application to WRAN signals.

*Cyclo-stationary Based Sensing*

Cyclo-stationary characteristics are present in a signal due to periodicities in signal or in its statistics like mean and autocorrelation. Such periodicities can also be introduced intentionally in the signal by means of modulation. The method of correlation generally derives the amount of similarities between two signals. It is very difficult to characterize random data like noise and thus it is devoid of periodicities. On the contrary the information signals possess such characteristics. The periodicities present an interesting way for user detection. Correlation of two information signals from primary users will result in maxima at certain values where the same procedure for random data will produce no such effect.

The cyclic spectral density (CSD) of the received signal is calculated as

$$S(f,\alpha) = \int_{\tau=-\infty}^{\infty} R_x^\alpha(\tau) e^{-j2\pi f \tau} d\tau$$

where

$$R_y^\alpha(\tau) = E\left[ y(n+\tau) y^*(n-\tau) e^{j2\pi\alpha n} \right]$$

is the cyclic autocorrelation function (CAF), α is the cyclic frequency, y(n) is the received signal, τ is the delay [7]. The computation of the CSD gives rise to peaks at the cyclic frequencies when they are equal to the fundamental frequency of the signal. Thus in the method of detection simply the cyclic frequencies are searched for the peaks. The detection of the peaks indicates the presence of primary user. The detection scheme is presented in Figure-2 wherein the received signal is shifted by α/2 and –α/2 in time domain. It is then passed through a band-pass filter thus narrowing band to frequency of interest. Conjugation of one of the terms is done and then both are multiplied followed by FFT operation to give the cyclic spectral density.

The Cyclo-stationary energy detector works well for low SNR values but since the spectrum is to be calculated for all the cyclic frequencies so the computation is large. It has an additional capability to distinguish between noise and primary user signal from the presence or absence of peaks at the cyclic frequencies.

In order to detect the WRAN signals, the DVB-T signal whose bandwidth is at 8 MHz is generated using the specifications given in the [8] . The signal is then applied to the Cyclo-Stationary detector.

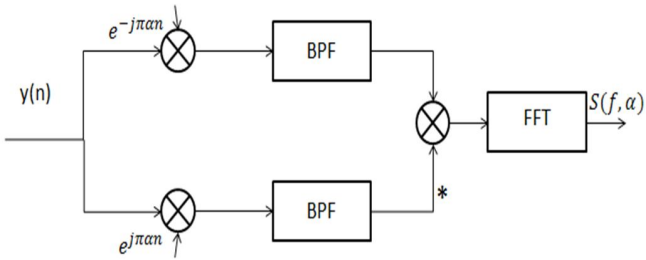

Fig.1 The Principle of Cyclo-Stationary Detector.

It outputs the cyclic spectral density (CSD) of the input signal. The CSD obtained is searched for peaks at the cyclic frequencies. The peak values are compared with a threshold [9] and if they are greater than the threshold, a primary user is confirmed [10].

## IV. ALGORITHM FOR OPTIMUM NUMBER OF USERS

The number of users can be varied according to the value of false alarm probability by finding a value of 'n' which minimizes the error ($P_f+P_m$). In order to minimize the error the method given in [11] is followed as

$$E = P_f + P_m$$

$$= 1 + \sum_{l=n}^{k} \binom{k}{l} \left[ p_f^l (1-p_f)^{k-l} - p_d^l (1-p_d)^{k-l} \right]$$

$$Let, G(n) = \sum_{l=n}^{k} \binom{k}{l} \left[ p_f^l (1-p_f)^{k-l} - p_d^l (1-p_d)^{k-l} \right]$$

$$\frac{\delta G(n)}{\delta n} \approx G(n+1) - G(n)$$

Now the optimal value of n is obtained when

$$\frac{\delta G(n)}{\delta n} = \binom{k}{n} \left[ (1-p_m)^n p_m^{k-n} - p_f^n (1-p_f)^{k-n} \right] = 0$$

$$\Rightarrow \left( \frac{p_f}{1-p_m} \right)^n = \left( \frac{p_m}{1-p_f} \right)^{k-n}$$

$$\Rightarrow \frac{k-n}{n} = \frac{\ln\left(\frac{p_f}{1-p_m}\right)}{\ln\left(\frac{p_m}{1-p_f}\right)} = \alpha (say)$$

$$n \approx \left\lceil \frac{k}{1+\alpha} \right\rceil$$

Thus optimal 'n' is given as

$$n_{opt} = \min\left( k, \left\lceil \frac{k}{1+\alpha} \right\rceil \right)$$

where ⌈ ⌉ denotes the ceiling function.

## V. THRESHOLD ADAPTATION

The threshold can be dynamically adapted by using the particle swarm optimization [12]. The technique is inspired by social behavior of bird flocking and does not require the objective function to be differentiable as required by classic optimization methods. In this technique particle change their positions based on the knowledge of themselves and others particles in the vicinity or swarm. In such a manner the particles update their position and move towards an optimum value.

The technique can be applied for threshold adaptation by replacing the positions of the particles with the threshold value of the secondary users. The threshold value decided by swarm in previous step ($G_{best}$) and the threshold value found out by the secondary users in previous step ($P_{best}$) are taken under consideration. In order to initialize the threshold of the secondary users, the threshold found out by them at the beginning are treated as initial values. The update equation for the optimization technique [13] is given as

$$V = C_0 \lambda(n) + C_1 r_1 \times (pbest - \lambda(n)) + C_2 r_2 \times (gbest - \lambda(n))$$

$$\therefore \lambda(n+1) = \lambda(n) + V$$

where '$\lambda$' is the threshold, '$C_0, C_1, C_2$' are constants and 'r1,r2' are random numbers between '0'and '1'. After updating threshold of each users, the minimum threshold of the swarm is finally chosen and the entire procedure is again repeated.

The iterative procedure is continued till a stopping condition is reached. The condition can be maximum number of iteration or difference between two thresholds values be less than a predefined tolerance.

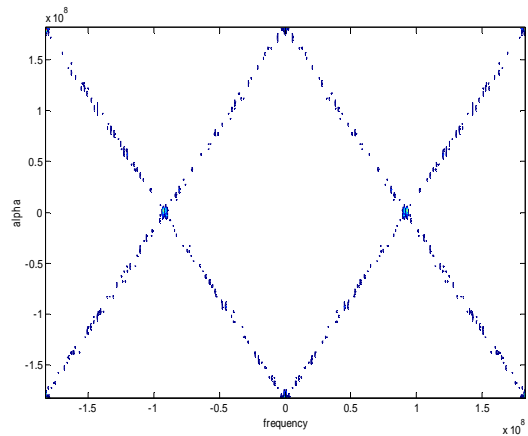

Fig. 2 Contour Diagram of WRAN CSD for SNR= -5 dB.

The condition can be stated as

$$|gbest(k+1) - gbest(k)| < \zeta$$

where $'\zeta'$ is the tolerance value.

## VI. RESULTS

DVB-T signal is generated from the equation given in the IEEE 802.22 standard specified the working group [8]. The carrier frequency is taken to be 91.44 MHz which is in the VHF range and is commonly used in TV broadcasting. The DVB-T signal is applied to Cyclo-Stationary detector which then determines the cyclic spectral density (CSD) for the signal. The CSD obtained is searched for peaks at frequencies given as

$$1) \alpha = 0 \;\&\; f = \pm f_c$$
$$2) \alpha = \pm 2f_c \;\&\; f = 0$$

Figure-2 shows the contour diagram of the CSD obtained from the detector with SNR= -5 dB.

The CSD is searched for peaks at the above mentioned cyclic frequencies and the magnitude at the required points is compared with a threshold. If the magnitude is greater than primary user is conformed. The procedure is repeated in case of number of users and different fusion rules are applied to find the Receiver Operating Characteristics (ROC) which is nothing but the plot of '$P_d$' vs '$P_f$'. Figure-3 gives the plot for different fusion rules in case of co-operative spectrum sensing at SNR=-5 dB and maximum number of users is 8.

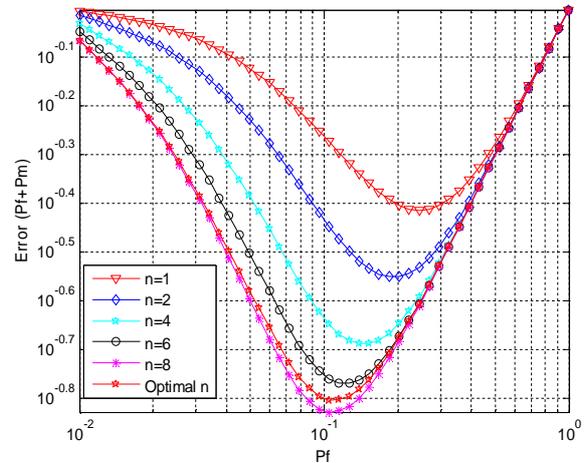

Fig. 3 Error vs Pf plot for SNR= -5 dB and OR fusion scheme.

Figure-4 shows the plot of Error '$P_m+P_f$' vs '$P_f$' for different fusion rules at SNR=-5 dB. It is clear from the figure that error for 'Optimal n' starts with coinciding with 'n=8' curve for low '$P_f$' values. As the value of the '$P_f$' is increased the optimal curve deviates from the 'n=8' curve and slowly goes on to 'n=6' curve. Thus as false alarm probability ($P_f$) increases the minimum number of users required for satisfactory performance reduces as indicated by the plot. Figure-5 shows optimal number of user's vs '$P_f$' plot for different SNR values.

Figure-6 displays the ROC curve after applying the particle swarm optimization technique. While applying the technique the constant were '$C_0=0$, $C_1=1$, $C_2=1$' where as the random values taken were 'r1=0.3811 and r2=0.1895'. Instead of having a tolerance value the procedure was repeated for a fixed number of iterations. The red curve shows the detection probability of a single user after adapting threshold using particle swarm optimization. When compared to the un-optimized curve it shows a significant improvement in detection.

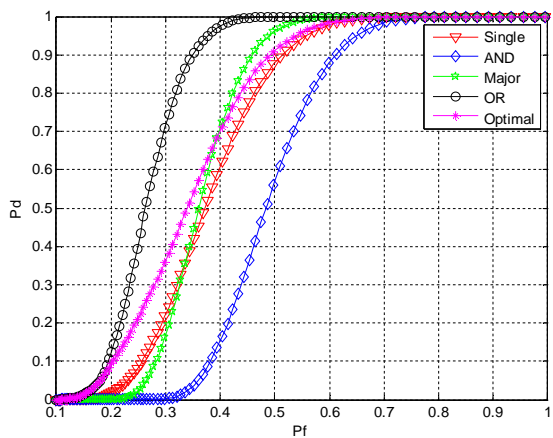

Fig. 4 ROC for different Fusion schemes with maximum number of users=8 and SNR= -5 dB.

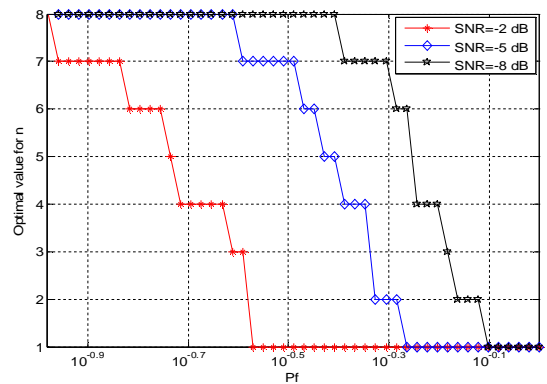

Fig. 5 Optimal n vs Pf plot for different SNR's and maximum number of users=8.

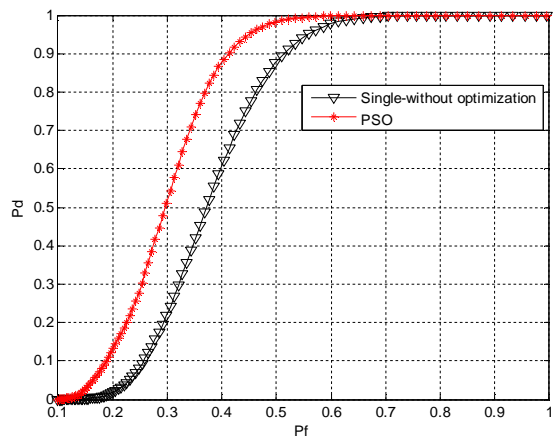

Fig. 6 ROC curve for PSO optimized and un-optimized threshold for single user.

## VII. CONCLUSION

In this paper, application of cyclo-stationary detector to WRAN signals under low SNR conditions has been simulated and detection is made possible by checking the cyclic frequencies. Numerical results show that optimal scheme varies the number of users so that error is kept as minimum as possible without compromising the detection probability. With the increase in false alarm probability the minimum number of users required for satisfactory performance decrease. Thus instead of keeping fixed number of users for fusion schemes in co-operative spectrum sensing, the number can be varied in accordance with the false alarm probabilities. Particle swarm optimization technique highlights a very good method which results in increased detection probability without putting many constraints on the objective function. Thus it allows it application even though the function is not differentiable as it is different from classic methods in terms of implementation.